\documentclass[prl,twocolumn]{revtex4-2}
\usepackage{amsmath}
\usepackage{graphicx}
\usepackage{color}
\usepackage[a-1b]{pdfx}
\newcommand{\gs}[1]{\textcolor{blue}{GS: \textbf{\textit{#1}}}}

\begin{document}
\title{Scale-free correlations in extremely persistent active matter}
\author{Grzegorz Szamel}
\affiliation{Department of Chemistry,
  Colorado State University, Fort Collins, CO 80523}

\begin{abstract}
  Scale-free correlations are believed to be generic in out of equilibrium systems but their explicit
  derivations are rare and often involve approximations. Here we analyze a system of infinitely persistent
  active Brownian particles in the low density and small-swim-velocity limit.
  We show that both the density and swim velocity  
  correlations decay algebraically as $r^{-(d+1)}$ (with logarithmic corrections in $d=2$).
  In reciprocal space these correlations manifest themselves as cusps at $k=0$.  The analytical
  predictions are confirmed by numerical simulations. The results
  were obtained through interaction with Claude (Opus 4.8 and 5, and Fable 5) and verified by the author.
\end{abstract}

\maketitle

\textit{Introduction. --}
It is often stated that in the absence of equilibrium, many correlation functions become
scale-free, \textit{i.e.}, they decay algebraically \cite{Schmitz1988,Grinstein, DorfmanARPC}. 
Standard examples include fluids in temperature gradient \cite{Kirkpatrick1982,Ronis1982} and sheared fluids
\cite{Lutsko2002}. In these systems
the drive, which generates non-equilibrium, is applied externally and at the macroscopic level.
Recently, there has been a lot of interest in a different class of non-equilibrium assemblies, active matter systems,
which are driven internally,
at the level of individual particles \cite{Ramaswamyrev1,Bechingerrev,Ramaswamyrev2,Marchettirev2}.
Thus, a natural question arises whether scale-free correlations can be found in active matter.
Two recent examples suggest that the answer, at least in some cases, is positive. 

First, Dasgupta \textit{et al.} \cite{Dutta} simulated dense systems of self-propelled particles and found
that in the limit of infinite persistence time of the self-propulsion
the length scales of correlations of the velocity and the self-propulsion force grow with the system
size.  Their findings fit nicely with the earlier discovery of velocity correlations in systems of
interacting active particles with finite persistence times \cite{Caprini2020,Henkes2020,Szamel2021},
where the velocity correlation length was found to grow proportionally to
the square root of the persistence time. Although this growth was only studied for a limited
range of persistence times, it suggested scale-free correlations in
systems with infinite persistence time, cut off by the system size for finite systems. 

Second, Damman \textit{et al.} \cite{Damman2024} investigated correlations in two-temperature mixtures,
which is an active matter model system qualitatively different from systems of interacting self-propelled particles.
They found scale-free correlations, decaying algebraically with  exponent $-2d$. These correlations
originate from triplets of particles at different temperatures. Very recently, Metzger \textit{et al.}
\cite{Metzger2026} attributed these scale free correlations to ``dangerously irrelevant nonlinearities'' in 
two-temperature mixtures. They showed that three-particle correlations also exhibit
power-law decay. 

Here we study correlations in systems of the kind considered by Dasgupta \textit{et al.},
\textit{i.e.} systems of interacting self-propelled particles, in the limit of infinite persistence time. 
To make an analytical study possible we focus on the low density and small self-propulsion limit.
We show that in this case the two-particle correlation function can be obtained analytically.
We find that correlations of particles' positions exhibit power-law decay with exponent $-(d+1)$,
with a logarithmic correction in $d=2$.
Furthermore, we show that swim velocity correlations also exhibit similar scale-free correlations. This
fact may sound peculiar since swim velocities are distributed independently, randomly,
and, in the infinite persistence time limit, do not evolve. However, scale-free swim velocity correlations reflect the
fact that particles positions adjust depending on their velocities and in the resulting stationary state
velocities and positions are strongly correlated. Both the positional and swim velocity correlations
in reciprocal space are finite in the zero wavevector limit. Their scale-free nature manifests itself by
cusps at the origin of reciprocal space. 
Our analytical predictions are verified by computer
simulations.

\textit{Model. --} We study a system of infinitely persistent active Brownian particles.
We focus on the $d=2$ dimensional case but we also mention some $d=3$ results.
Additional discussion of the $d=3$ case can be found in Appendix A. The particles
move under the combined influence of self-propulsion, interparticle interactions and thermal noise.
The equations of motion read
\begin{equation}\label{eq:eom}
  \gamma\dot{\mathbf{r}}_i = \gamma v_0 \mathbf{e}_i + \sum_j \mathbf{F}(\mathbf{r}_{ij}) +
  \boldsymbol{\xi}_i,
\end{equation}
where $\mathbf{r}_i$ is the position of particle $i$, $\gamma$ is the friction coefficient,
$v_0$ is the swim velocity, $\mathbf{e}_i$ is a unit vector with a random orientation representing
the direction of the self-propulsion of particle $i$, 
$\mathbf{F}(\mathbf{r}_{ij})$, where $\mathbf{r}_{ij} = \mathbf{r}_i-\mathbf{r}_j$, 
is the force on particle $i$ due to particle $j$ and
$\boldsymbol{\xi}_i$ represents thermal white Gaussian noise, with
$\left<\xi_{i\alpha}\xi_{j\beta}\right> = 2T\gamma\delta_{ij}\delta_{\alpha\beta}$, where Greek symbols
indicate Cartesian coordinates. We consider hard-sphere interactions, which implies that the force
in Eq. \eqref{eq:eom} needs to be replaced by a hard-core exclusion condition.

\textit{Pair distribution. --} To calculate the pair distribution  function $g(\mathbf{r};\mathbf{e}_1,\mathbf{e}_2)$
in the low-density limit we only need to consider the two-particle
dynamics. It is convenient to formulate the problem in terms of the differential equation for $g$,
\begin{equation}\label{eq:pair1}
  2D_0 \nabla^2 g 
  - \nabla\cdot \mathbf{u}g = 0 \qquad
  \mathbf{u}=v_0 (\mathbf{e}_1-\mathbf{e}_2).
\end{equation}
The hard-core exclusion condition is implemented as a no-flux boundary condition at touching, \textit{i.e} at
$r\equiv |\mathbf{r}| = \sigma$, where $\sigma$ is the hard sphere diameter,
\begin{equation}\label{eq:pair1bc}
\hat{\mathbf{r}}\cdot\!\left[-2D_0\nabla g+\mathbf{u}\,g\right]_{r=\sigma}=0,
\end{equation}
where $\hat{\mathbf{r}}=\mathbf{r}/r$. With no self-propulsion, $g=1$ for $r\ge 1$.

We focus on the change of the pair distribution due to the self-propulsion, $\delta g = g -1$,
which depends on particle separation $\mathbf{r}$ and relative drift $\mathbf{u}$. 
We solve for $\delta g$ adopting the
method used by B\l awzdziewicz and Szamel \cite{BS1993} to find the low density pair distribution
function of a sheared hard sphere suspension.
Specifically, we expand $\delta g$ in terms of solutions of Eq. \eqref{eq:pair1}. In $d=2$, it is convenient to
use the following solutions as the expansion basis,
\begin{equation}\label{eq:basis}
  T_n(\mathbf{r},\mathbf{u}) = 
  e^{\boldsymbol{\kappa}\cdot\mathbf{r} } K_n(\kappa r) \cos(n\theta),
\end{equation}
where $\boldsymbol{\kappa}=\mathbf{u}/(4D_0)$,
$K_n$ is the modified Bessel function of the second kind, $\kappa=|\boldsymbol{\kappa}|$, and
$\theta$ is the angle between $\mathbf{r}$ and $\mathbf{u}$. We note that $(4\pi D_0)^{-1}T_0$ is the Green's
function of Eq. \eqref{eq:pair1}. We expand $\delta g$ as $\delta g = \sum_n c_n T_n$.
The no-flux boundary condition \eqref{eq:pair1bc} can then be used to derive a system of linear
equations for expension coefficients $c_n$. In $d=2$ we get 
\begin{equation}\label{eq:lineqs}
  \sum_n M_{mn} c_n = -2\pi \delta_{m,1},
\end{equation}
where coefficients $M_{mn}$ can be expressed in terms of Bessel functions at argument $\kappa\sigma$.
It can be shown analytically that the coefficients in the first row, $M_{0n}$, do not depend on $n$,
which implies that $\sum_n c_n =0$. Physically, this relation expresses the fact that the source (the right-hand-side)
in the equation for $\delta g$ is a pure dipole and as a result the flux of $\delta g$ through any
disk centered at the origin vanishes. 

As in Ref. \cite{BS1993}, for a finite swim velocity
one could find out numerically how many basis functions \eqref{eq:basis} are needed to get convergent results.
In this way pair distribution distortion $\delta g$ can be found numerically exactly. In the following we
focus on the small-swim-velocity limit, which can be discussed fully analytically. 

In the small-swim-velocity limit, $\kappa\sigma\ll 1$, the analysis of coefficients $M_{mn}$ shows
that only $T_0$ and $T_1$ contribute and $\delta g$ has the following form,
\begin{equation}\label{eq:pair2}
  \delta g(\mathbf{r},\mathbf{u})
  = 2 (\kappa\sigma)^2 \left(T_0(\mathbf{r},\mathbf{u}) - T_1(\mathbf{r},\mathbf{u}) \right). 
\end{equation}
Physically, Eq. \eqref{eq:pair2} can be obtained by noting that the small $\kappa\sigma$,
perturbative solution of Eq. \eqref{eq:pair1} can be matched with 
the small $r$ form of $T_1$, which fixes $c_1$, and then using the no flux condition $\sum_n c_n =0$
to get $c_0$.

\begin{figure}[t]
  \centerline{\includegraphics[width=\columnwidth]{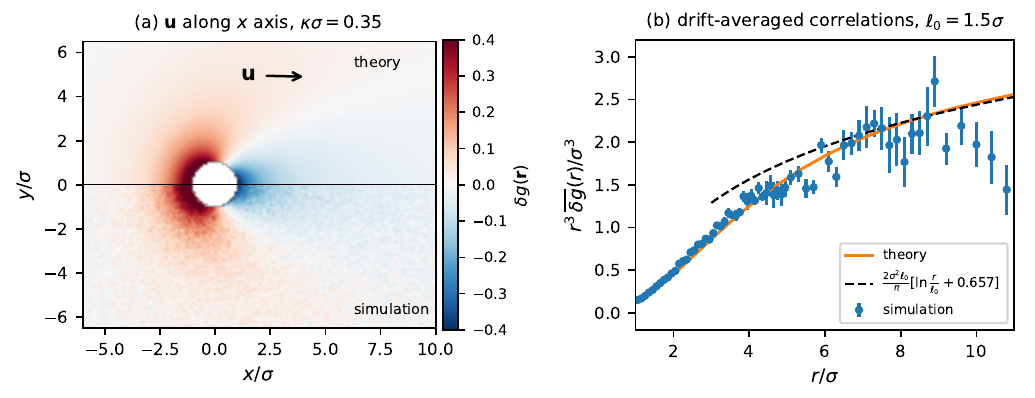}}
  \caption{(a) Pair distribution function distortion, $\delta g(\mathbf{r};\mathbf{e}_1,\mathbf{e}_2)$,
    in $d=2$ and for
    relative drift $\mathbf{u}\propto\mathbf{e}_1-\mathbf{e}_2$ along $x$ axis for
    $\kappa\sigma=0.35$. Upper half shows theoretical expression \eqref{eq:pair2} and lower half shows
    results of Langevin simulations of the relative motion. (b) Drift-averaged $d=2$ distortion compensated
    by $\left(r/\sigma\right)^3$, $r^3 \overline{\delta g}(r)/\sigma^3$: full theory (solid line), a
    large $r$ asymptotics \eqref{eq:tail1} (dashed line) and Langevin simulations (symbols).}
  \label{fig1}
\end{figure}

Solution \eqref{eq:pair2} is shown in the upper half of Fig. \ref{fig1}(a). It exhibits an anisotropic ``wake'',
with an upstream accumulation and a depleted downstream region decaying as $r^{-3/2}$ along the
$\mathbf{u}$ axis. It agrees well with the results of computer simulations of the Langevin equation
describing the relative motion of two particles, \textit{i.e.} the Langeving equation corresponding
to differential equation \eqref{eq:pair1} with boundary condition \eqref{eq:pair1bc}, which are shown
in the lower half of Fig. \ref{fig1}(a).

\textit{Scale-free correlations.--} To analyze the distribution of the positions, \textit{i.e.}
$\delta g(\mathbf{r},\mathbf{u})$ averaged over relative drift, $\overline{\delta g}$, it is convenient
to start from the Fourier transform of solution \eqref{eq:pair2},
\begin{equation}\label{eq:pair3}
  \delta g(\mathbf{k},\mathbf{u})
  = 2\pi\sigma^2\left(1-\frac{k^2}{k^2+i\mathbf{u}\cdot\mathbf{k}/(2D_0)}\right).
  \end{equation}
Averaging $\delta g(\mathbf{k},\mathbf{u})$ over the distribution of the relative drift,
$p(\mathbf{u}) = \left(\pi^2\sqrt{4v_0^2-u^2}\right)^{-1}$ we get
\begin{equation} \label{eq:pairk1}
\overline{\delta g}(k)=2\pi\sigma^2\left[1-\frac{2q}{\pi\sqrt{q^2+1}} 
K\left(\frac{1}{\sqrt{q^2+1}}\right)\right],
\end{equation}
where $q=kD_0/v_0$ and $K(m)=\int_0^{\pi/2}(1-m^2\sin^2\theta)^{-1/2} d\theta$ is the complete elliptic
integral of the first kind \cite{Abramowitz}. In the $k\to 0$ limit expression \eqref{eq:pairk1} exhibits
a logarithmically enhanced cusp, $\overline{\delta g}(k)\propto q\ln q$, which is reciprocal
space signature of power-law decaying correlations in the real space. Indeed, in real space in the large $r$
limit $r\gg l_0\equiv D_0/v_0$ we get
\begin{equation}\label{eq:tail1}
\overline{\delta g}(r) \approx \frac{2\sigma^2\ell_0}{\pi r^3}\left[\ln\frac{r}{\ell_0}+4 C_0\right],
\end{equation}
where $C_0$ is given by a one-dimensional integral involving Bessel functions;
numerically $C_0\approx 0.164$.
The real-space correlations are scale-free; for $r\gg l_0$ there is no further length in the problem. Fig. \ref{fig1}(b)
shows the comparison of the complete theoretical expression, the large $r$ asymptotics \eqref{eq:tail1}
and Langevin simulations. 

We note that Eq. \eqref{eq:pair3} is valid for any $d$ if $\sigma^2$ is replaced by $\sigma^d$.
In $d=3$ the drift distribution is different than in $d=2$ and the drift-averaged distortion in
reciprocal space reads
$\overline{\delta g}(k)=2\pi\sigma^3[1-2q\arctan(1/q)+ q^2\ln(1+1/q^2)]$.
In real space, in $d=3$ we get a pure power-law tail,
$\overline{\delta g}(r) \approx 2\sigma^3l_0/r^4$.

Physically, power-law decay of drift-averaged correlations originates from the power-law
decay of the downstream distortion. Isotropic average converts the ensemble of downstream
distortions into the isotropic power law \eqref{eq:tail1}. The logarithmic enhancement in $d=2$
originates from the finite probability of nearly parallel swimmers, $\lim_{u\to 0} p(\mathbf{u})>0$. 

\textit{Swim velocity correlations.--} The natural observables of persistent motion are the equal-time
swim velocity correlation functions \cite{Szamel2021} 
$\omega_\parallel(k)=N^{-1}\langle|\sum_j(\hat{\mathbf{k}}\cdot\mathbf v_j)
e^{-i\mathbf{k}\cdot\mathbf{r}_j}|^2\rangle$
and its transverse analog, where $\hat{\mathbf{k}}=\mathbf{k}/k$ and
swim velocities are $\mathbf{v}_j=v_0\mathbf{e}_j$. Separating the self part we write 
$\omega_\parallel(k) = v_0^2/2 + \Omega_\parallel(k)$ and similarly for the transverse correlations.
In the low-density limit the distinct correlations can be calculated from $\delta g(\mathbf{k},\mathbf{u})$.

In the small-swim-velocity limit the distortion depends on the orientations only through
$\mathbf{k}\cdot\left(\mathbf{e}_1-\mathbf{e}_2\right)$ and thus the distinct transverse correlations 
vanish at leading order, $\Omega_\perp(k)=0$.

For the distinct longitudinal correlations we get
\begin{align}\label{eq:pairvelpar}
\Omega_\parallel(k)&=-2\pi\sigma^2\rho v_0^2\,C_\parallel(q), \\ \nonumber
C_\parallel(q)&=\frac{2q}{\pi\sqrt{q^2+1}}
\left[2(q^2+1)\left(K(m)-E(m)\right)-K(m)\right],
\end{align}
where $E(m)=\int_0^{\pi/2}(1-m^2\sin^2\theta)^{1/2}d\theta$ is the complete elliptic
integral of the second kind, and
$m=1/\sqrt{q^2+1}$. In $d=3$ in Eq. \eqref{eq:pairvelpar} $\sigma^2$ is replaced by $\sigma^3$ and
the corresponding kernel reads 
$C_\parallel=\frac{1}{3}[2q\arctan(1/q)+2q^2-(3q^2+2q^4)\ln(1+1/q^2)]$.

\begin{figure}[t]
  \centerline{\includegraphics[width=\columnwidth]{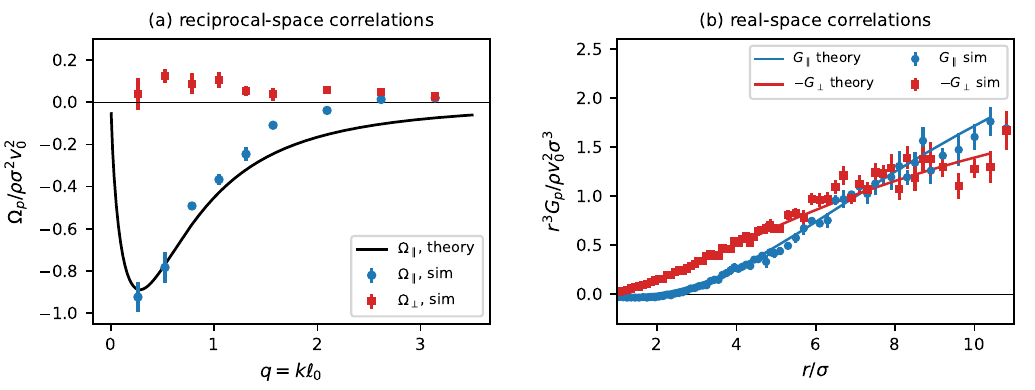}}
  \caption{(a) Swim velocity correlations in $d=2$ at $l_0=1.5\sigma$. (a) Reciprocal-space correlations 
    $\Omega_\parallel$ and $\Omega_\perp$: full theory (line) and Langevin simulations (symbols). 
  (b) Real-space correlations, Eq. \eqref{eq:Gten1}, compensated by $r^3$: full theory (lines) and
    Langevin simulations (symbols).}
  \label{fig2}
\end{figure}

The $d=2$ result, Eq. \eqref{eq:pairvelpar} is shown in Fig. \ref{fig2}(a). It develops a dip at $kl_0\approx 0.28$ 
and returns to its asymptotic value as $k^{-2}$. More interestingly, it approaches its $k=0$ limit non-analytically,
\begin{equation}\label{eq:pairvelparsmallk}
\Omega_\parallel(k)=-4\rho\sigma^2v_0^2\,\ell_0k\left[\ln\frac{4}{\ell_0k}-2\right]
+O\left(k^3\ln k\right)
\end{equation}
[in $d=3$, as $k\to 0$, $\Omega_\parallel(k)=-(2\pi^2/3)\rho\sigma^3v_0^2\,\ell_0k+O(k^2\ln k)$]. 
The cusp at $k=0$ is the reciprocal space signature of power-law correlations in real space, where
the object of interest is the distinct swim velocity correlation tensor,
\begin{equation}\label{eq:Gdef}
G_{\alpha\beta}(\mathbf{r})=\rho v_0^2\langle e_{1\alpha}e_{2\beta}\delta g(\mathbf{r},\mathbf{u})\rangle,
\end{equation}
where $\langle \ldots \rangle$ denotes the orientation average over $\mathbf{e}_i$, i=1,2.
The isotropy dictates that $G_{\alpha\beta}(\mathbf{r})=G_\parallel(r)\hat{r}_\alpha\hat{r}_\beta
+G_\perp(r)(\delta_{\alpha\beta}-\hat{r}_\alpha\hat{r}_\beta)$ with
\begin{equation}\label{eq:Gten1} 
  G_\parallel=\rho v_0^2\langle(\hat{\mathbf{r}}\cdot\mathbf{e}_1)(\hat{\mathbf{r}}\cdot\mathbf{e}_2)
  \delta g\rangle,
\quad
G_\perp=\rho v_0^2 \langle(\hat{\mathbf{t}}\cdot\mathbf{e}_1)(\hat{\mathbf{t}}\cdot\mathbf{e}_2)
\delta g \rangle,
\end{equation}
with $\hat{\mathbf{t}}\perp\hat{\mathbf{r}}$.
Velocity correlations are scale-free with the same exponents as density correlations,
\begin{equation}\label{eq:Gten2}
  G_{\alpha\beta}(\mathbf{r})\approx \frac{2\rho v_0^2\sigma^2 l_0}{\pi}
  \left(3\hat{r}_\alpha\hat{r}_\beta-\delta_{\alpha\beta}\right)\frac{\ln(r/l_0)}{r^3} \text{ for } r \gg l_0,
\end{equation}
[in $d=3$, $G_{\alpha\beta} \approx
  (2\rho v_0^2\sigma^3 l_0/3)(4\hat{r}_\alpha\hat{r}_\beta-\delta_{\alpha\beta})/r^4$].
As shown in Fig. \ref{fig2}(b), swim velocities are correlated along the separation and anticorrelated transverse to it.

\begin{figure}[t]
\centerline{
\includegraphics[width=0.9\columnwidth]{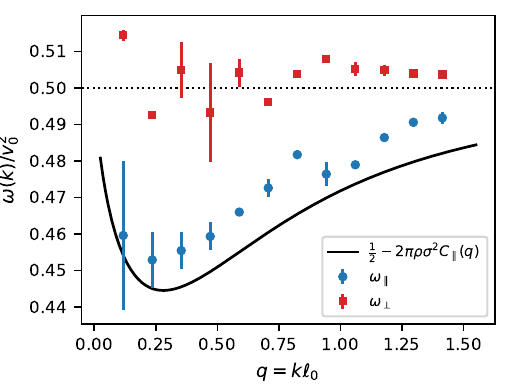}}
\caption{Longitudinal, $\omega_\parallel$, and transverse, $\omega_\perp$, velocity
  correlations in reciprocal space. Number density $\rho\sigma^2=0.0623$, $l_0=1.5\sigma$.
  Theory (full line for $\omega_\parallel$ and horizontal dashed line for $\omega_\perp$) compared
  with many-particle siulations (symbols).
}
\label{fig3}
\end{figure}

Finally, in Fig. \ref{fig3} we compare the complete longitudinal and transverse correlation functions,
$\omega_\parallel$ and $\omega_\perp$, with many-particle simulations at a small but finite density.
We observe a dip in $\omega_\parallel$ of about 11.0\% that agrees well with the simulated dip of
about 9.4\% and wavevector independence of $\omega_\perp$ to within 1\%.

\textit{Discussion and outlook.--}
A dilute system of infinitely persistent active hard spheres is an example of an active matter system
that can be analyzed numerically exactly. 
The analysis is free from linearization, weak-coupling expansion or field-theoretic approximations.
In the small-swim-velocity limit one can derive an explicit 
analytical expression for the distortion of the pair distribution due to the self-propulsion.
It exhibits scale-free, algebraically decaying with integrable exponent $-(d+1)$,
correlations of both positions and swim velocities of different particles. The integrability of these 
correlations leads to their finite values at the origin of reciprocal space with cusps at $k=0$
that are the reciprocal space counterparts of algebraic decays in real space.
Scale-free correlations originate from algebraically decaying depletion downstream from the relative
drift averaged over the drift orientation and magnitude. This mechanism is different from the triplet-induced
mechanism in two-temperature mixtures \cite{Damman2024,Metzger2026}.

The work presented here can be extended mirroring the very much analogous studies of dilute sheared colloidal
suspensions. For example, one can repeat the boundary-layer analysis of Dhont \cite{Dhont1989} or 
extend the small swim velocity analysis presented here to larger $v_0$ along the lines of Ref. \cite{BS1993}.
In addition, one can use the pair distribution distortion to calculate transport coefficients of dilute
active systems, following Bergenholtz \textit{et al.} \cite{Bergenholtz2002}. 
Finally, it would be interesting to go beyond the low-density limit, but very likely the analysis then 
becomes considerably more involved. 

\textit{Acknowledgments} --- This project benefitted from discussions at the
COMPLEXSYS26 program at the Kavli Institute for Theoretical Physics (KITP) on topics ranging from using
LLMs in research to scale-free correlations in active matter. I especially thank
F. Zamponi, P. Charbonneau and C. Nardini. This research was supported in part by NSF
Grant~PHY-2309135 to KITP. I also acknowledge the support of NSF Grant No.~CHE 2154241
and the use of Gemini AI for grammar checking.

\textit{Appendix A: $d=3$ basis functions and orientational averaging.--}
In $d=3$, we expand $\delta g$ as $\delta g = \sum_n c_n W_n$.
It is convenient to use the following solutions of Eq. \eqref{eq:pair1} as the expansion basis,
\begin{equation}\label{eq:basis3d}
  W_l(\mathbf{r},\mathbf{u}) = 
  e^{\boldsymbol{\kappa}\cdot\mathbf{r} } \tilde{k}_l(\kappa r) P_l(\theta),
\end{equation}
where $\tilde{k}_l$ are the modified spherical Bessel functions of the second kind,
$\tilde k_l(z)=\frac{e^{-z}}{z}\sum_{j=0}^l 
\frac{(l+j)!}{j!\,(\ell-j)!\,(2z)^{j}}$, which differ from the standard functions $k_l$ by a multiplicative
constant, $\tilde{k}_l(z) = \left(2/\pi\right) k_l(z)$, and $P_l$ is the Legendre polynomial. We note that
in $d=3$ the Green's function of Eq. \eqref{eq:pair1} is
$(\kappa/(8\pi D_0))W_0 = e^{\boldsymbol{\kappa}\cdot\mathbf{r}-\kappa r}/(8\pi D_0 r)$.

Again, since the source in Eq. \eqref{eq:pair1} is a pure dipole, the flux of $\delta g$ vanishes, 
which implies that $\sum_n c_n = 0$. Next, we note that if $c_1=-\kappa^3\sigma^3$, then the small
$\kappa\sigma$, perturbative solution of Eq. \eqref{eq:pair1} matches the small $r$ limit of $c_1 W_1$.
These two facts lead to the following form of the $d=3$ small swim velocity distortion 
\begin{equation}\label{eq:pair4}
  \delta g(\mathbf{r},\mathbf{u})
  = (\kappa\sigma)^3 \left(W_0(\mathbf{r},\mathbf{u}) - W_1(\mathbf{r},\mathbf{u}) \right). 
\end{equation}

In $d=3$, the distribution of the relative drift is $p(\mathbf{u}) = u/(8\pi v_0^2)$. Averaging
$\delta g(\mathbf{r},\mathbf{u})$ over this distribution one can derive a closed form expression
for $\overline{\delta g}(r)$. It exhibits a pure power-law decay, $\overline{\delta g}(r) \approx 2\sigma^3l_0/r^4$.

\textit{Appendix B: Computer simulations.--}
Analytical predictions were tested against computer simulations of the Langevin equation
corresponding to differential equation \eqref{eq:pair1} with boundary condition \eqref{eq:pair1bc}.
$M=3\times 10^5$ independent walkers were propagated in a box of size $L=32\sigma$. 
The boundary condition was implemented by the image reflection method, verified against the well-established
event-driven Brownian dynamics method of Scala \textit{et al.} \cite{Scala2007}.
Many-particle simulations evolved $N=400$ hard disks in a box of $L=80\sigma$ resulting in
dimensionless number density $\rho\sigma^2=0.0623$. The hard core exclusion condition
was implemented via pairwise image reflection.

\end{document}